# Energy spectrum design and potential function engineering

A. D. Alhaidari[a] and T. J. Taiwo[b]

[a] *Saudi Center for Theoretical Physics, P.O. Box 32741, Jeddah 21438, Saudi Arabia*

[b] *Physics Department, Untied Arab Emirate University, P.O. Box 15551, Al-Ain, United Arab Emirates*

**Abstract**: Starting with an orthogonal polynomial sequence $\{p_n(s)\}_{n=0}^\infty$ that has a discrete spectrum, we design an energy spectrum formula, $E_k = f(s_k)$, where $\{s_k\}$ is the finite or infinite discrete spectrum of the polynomial. Using a recent approach for doing quantum mechanics based, not on potential functions but, on orthogonal energy polynomials, we give a local numerical realization of the potential function associated with the chosen energy spectrum. In this work, we select the three-parameter continuous dual Hahn polynomial as an example. Exact analytic expressions are given for the corresponding bound states energy spectrum, scattering states phase shift, and wavefunctions. However, the potential function is obtained only numerically for a given set of physical parameters.

## 1. Introduction

The structure of a quantum mechanical system is defined by its discrete bound states and resonances whereas the dynamics is governed by its continuum scattering states. The discrete energy spectrum could be finite or countably infinite. If the system is modeled by a potential function, then there is a one-to-one correspondence between the complete energy spectrum (continuous + discrete) and the potential. The energy spectrum of a totally confined system consists only of discrete bound states (structure). On the other hand, dynamical information about a system interacting with its surroundings is found in the continuous part of the energy spectrum. Such information is contained in the scattering matrix (or scattering phase shift). All well-known quantum systems associated with exactly solvable potential functions (e.g., the harmonic oscillator, Coulomb, Morse, Pöschl-Teller, Eckart, Scarf, etc.) have simple discrete energy spectra $\{E_k\}$ that go like $k$, $(k+\mu)^2$ or $(k+\mu)^{-2}$, where $\mu$ is some dimensionless parameter. It would be interesting and fruitful to design systems with a richer energy spectrum (such as, $E_k \sim e^{-k^2}$, $E_k \sim \sinh(k)$, $E_k \sim \sin(k\pi/2N)$ with $k \leq N$, etc.) and to construct the corresponding potential function. In this work, we propose a method to accomplish just that. If successful, it could be considered as one of the solutions to the inverse problem. That is, constructing the potential function using knowledge of the energy spectrum data. As will be shown in the text, this solution is not unique due to an equivalence generated by similarity transformations of the corresponding Hamiltonian matrix. Now we make an indispensable digression to address the notion of a potential function in quantum mechanics.

The concept of a potential function originated long before the inception of quantum mechanics. It is rooted in our understanding of classical mechanics where the total energy of a particle (matter) moving in a field (non-matter) is the sum of its kinetic energy and *potential energy*. The potential energy changes with the *position* of the particle, hence the concept of a potential function. For example, the potential function of a massive particle moving in the gravitational field of a point mass $M$ at the origin is $MG/r$ ($G$ being the gravitational constant and $r$ the radial distance from the point mass) whereas for a particle of charge $q$ moving in the electric field of a point charge $Z$ at the origin it is $qZ/r$. Also, the potential function for a massive particle attached to a linear massless spring of constant $k$ is $\frac{1}{2}kx^2$, etc. Subsequently, different types of potential functions where proposed to describe certain aspects of the system in a complicated surroundings. Examples include the Yukawa potential $Qe^{-\lambda r}/r$ to describe screening of the inverse square force, the oscillator potential $\omega^2 r^2$ to describe confined oscillatory motion, the Morse potential $D(e^{-2\lambda x} - 2\mu e^{-\lambda x})$ to describe the vibrations in a diatomic molecule, etc. The potential function concept was then carried over to quantum mechanics, through a particular construction of the system's Hamiltonian, despite the fact that none of the postulates of the theory requires it. In fact, in the postulates of quantum mechanics there are only two fundamental objects: the space-time wavefunction $\Psi(t,\vec{r})$ and the Hamiltonian operator $H$. Other operators are developed to correspond to the desired physical property to be measured (e.g., position, linear momentum, angular momentum, spin, etc.). The wavefunction allows for an evaluation of the expectation values (measurements) of physical observables at a given time and the Hamiltonian operator determines their time development. It is by (arbitrary) choice that the Hamiltonian was split into the sum of a kinetic energy operator and a potential function as $H = T + V$; a suggestion that was inspired by, and carried over from, classical mechanics, which over many decades (of success) became the tradition. An undergraduate student in quantum mechanics is taught to make a one-to-one correspondence between a given physical system and its classical potential function model. In fact, the present undergraduate physics curricula make classical mechanics a prerequisite to quantum mechanics courses. Hence, whatever one does or says to describe the quantum mechanical problem, the student will always ask about the potential function that he/she *needs* to incorporate into the wave equation so that he/she can solve that problem. Imagine describing the oscillator problem by explaining in full details how the particle moves and then ask the students "what is the energy spectrum?" Admittedly, this is not an easy task for the students because they will have to start by solving the inverse problem: finding the potential function from the description given then substituting that into the Schrödinger equation to solve for the energy spectrum. It is that complicated! They were taught this way - the only way: They *need* the potential function to proceed. It is more than just a tradition - it is a religion. Moreover, exact solvability of the wave equation in the potential function picture limits the number of analytically realizable quantum systems. These systems are well known and arranged into a small number of classes. Each class is associated with a given potential function like the Coulomb, harmonic oscillator, Morse, Eckart, etc.

Nevertheless, an alternative formulation of doing quantum mechanics was developed recently [1-3] on the premise that the set of analytically realizable quantum systems is much larger than the set of exact solutions of the Schrödinger equation in the potential function picture. Equivalently, we are affirming that the representation of the Hamiltonian operator in the wave equation, $i\hbar \frac{\partial}{\partial t}\Psi = H\Psi$, as the sum $H = T + V$ is a very restrictive choice that limits the number of analytically realizable physical systems. In this alternative formulation, the theory of orthogonal polynomials and special functions plays a major role. We believe that the potential function picture in quantum physics is a limitation. The objective of the alternative formulation is to obtain a set of analytically realizable systems, which is much larger than the set in the conventional potential formulation. Sometimes, this implies that the potential functions corresponding to the newly found systems may not have analytic realizations or that the associated wave equation cannot be written in the conventional format (that is, it could become a differential equation of order higher than two or with nonlocal potential, etc.). In the absence of a potential function, the new formulation gives the wavefunction as pointwise convergent series of a complete set of square-integrable functions in configuration space. The expansion coefficients of the series are orthogonal polynomials in the energy and/or physical parameters. These energy polynomials carry all of the physical information about the system. Nonetheless, to establish a correspondence between the new formulation and the potential formulation, procedures were established to obtain accurate enough numerical representations of the potential functions of the new quantum systems for a given set of physical parameters [4]. These procedures were used successfully in several studies [4-7] to derive novel quantum systems and obtain their corresponding potential functions.

In this work, we present a general scheme for obtaining a class of quantum systems associated with the three-parameter continuous dual Hahn orthogonal polynomials $S_n^\mu(z^2;a,b)$ for a given choice of energy spectrum formula $E(z^2)$ and a given set of basis functions in configuration space, $\{\phi_n(x)\}$. The energy spectrum, scattering phase shift and wavefunction are obtained *analytically*. However, the corresponding potential function can only be derived *numerically* for a given set of physical parameters. By choosing the energy as a special function of the polynomial argument, $E(z^2)$, we show in Section 2 how the energy spectrum, scattering phase shift and wavefunction can be derived analytically. In section 3, we show how to obtain the matrix elements of the potential function for a given choice of square-integrable basis functions $\{\phi_n(x)\}$. Finally in Section 4, we make local plots of the potential function for illustrative examples and for a given set of physical parameters.

## 2. The quantum system

We start by giving a brief account of the formulation of quantum mechanics based on orthogonal energy polynomials but with no mention at all of any potential function. In the atomic units $\hbar = M = 1$ and in one dimension, the total wavefunction in this formulation is written as the following Fourier expansion in the energy [1-3]

$$\Psi(x,t) = \int_\mathcal{X} e^{-iEt}\psi(x,E)dE + \sum_{k=0}^{N} e^{-iE_k t}\psi_k(x), \qquad (1)$$

where $\mathcal{X}$ is the continuous energy interval/intervals and $\{E_k\}$ are the discrete bound state energies. The continuous and discrete Fourier components $\psi(x,E)$ and $\psi_k(x)$ are written as the following pointwise convergent series

$$\psi(x,E) = \sqrt{\rho(z)} \sum_{n=0}^{\infty} P_n(z^2)\phi_n(x), \qquad (2a)$$

$$\psi_k(x) = \sqrt{\omega_k} \sum_{n=0}^{\infty} P_n(z_k^2)\phi_n(x), \qquad (2b)$$

where $z^2$ is some proper function of the energy, $z^2(E)$,[†] and $\{\phi_n(x)\}$ is a complete set of square-integrable functions. Moreover, $P_n(z^2)$ is a polynomial in $z^2$ of degree $n$ satisfying the following general orthogonality and recursion relation [1-3]

$$\int_{\mathcal{X}} \rho(z) P_n(z^2) P_m(z^2) dz + \sum_k \omega_k P_n(z_k^2) P_m(z_k^2) = \delta_{n,m}, \tag{3a}$$

$$z^2 P_n(z^2) = A_n P_n(z^2) + B_{n-1} P_{n-1}(z^2) + B_n P_{n+1}(z^2). \tag{3b}$$

where $A_n$ and $B_n$ are real numbers that are independent of $z$ such that $B_n \neq 0$ for all $n$. The three-term recursion relation (3b) gives all $\{P_n(z^2)\}$ starting with the two initial values $P_0(z^2) = 1$ and $P_1(z^2) = (z^2 - A_0)/B_0$. The wavefunction (2a) associated with the continuous spectrum is characterized by *bounded* oscillations that do not vanish all the way to the boundaries of space. However, the wavefunction (2b) associated with bound states is characterized by a finite number of oscillatory-like behavior (with a number of nodes that equals the bound state excitation level) that vanishes rapidly at the boundaries. Attempting to evaluate the wavefunction at an energy that does not belong to the continuous or discrete spectrum will only result in a diverging series (2). That is, the result is non-stable endless oscillations that grows without bound all over space as the number of terms in the sum (2) increases.

In addition to the orthogonality property (3a) and recursion relation (3b), polynomials that are compatible with this alternative formulation of quantum mechanics must have an oscillatory asymptotics ($n \to \infty$) that takes the following form

$$\lim_{n \to \infty} p_n(s) = \frac{1}{n^\tau \sqrt{\rho(s)}} \cos\left[n^\sigma \varphi(s) + \delta(s)\right], \tag{4}$$

where $\tau$ and $\sigma$ are positive real parameters, $\varphi(s)$ is an entire function, and $\delta(s)$ is the scattering phase shift. If $\sigma \to 0$ then $n^\sigma \to \ln(n)$. Almost all known hypergeometric orthogonal polynomials that populate the physics and engineering literature [8] have this asymptotic property. This includes the Wilson, continuous Hahn, continuous dual Hahn, Meixner-Pollaczek, Jacobi, Laguerre, Gegenbauer, Chebyshev, Hermite, etc. Bound states (if they exist) occur at a finite or countably infinite set $\{s_k\}$ that makes the scattering amplitude [left factor of the sinusoidal in (4)] vanish. That is, where $\rho(s)|_{s \to s_k} \propto \delta(s - s_k)$. Some of the polynomials mentioned are not endowed with a discrete spectrum that could correspond to bound states (e.g., the Hermite, Chebyshev, Laguerre, Jacobi, etc.). For a detailed discussion about the connection between the asymptotics of such polynomials and scattering, one may consult [9-11] and references cited therein.

For the current problem, we choose the orthogonal polynomial $P_n(z^2)$ to be a two-parameter special case of the continuous dual Hahn polynomial: $P_n(z^2) = S_n^\mu(z^2; a, a)$. The definition of the orthonormal version of the polynomial, $S_n^\mu(z^2; a, b)$, and its properties that are relevant to our study are given in Section IV.B and in the Appendix of Ref. [1]. Consequently, if $\mu < 0$ then in addition to the continuous spectrum, the system will have a finite number of bound states with an energy spectrum that could be obtained by solving for $\{E_k\}$ in the following energy spectrum formula [1]

$$z^2(E_k) = -(k+\mu)^2, \tag{5}$$

where $k = 0,1,...N$ and $N$ is the largest integer less than or equal to $-\mu$. The scattering phase shift associated with the continuous spectrum is obtained from the asymptotics ($n \to \infty$) of $S_n^\mu(z^2;a,a)$ as [1]

$$\delta(E) = \arg\Gamma[2iz(E)] - \arg\Gamma[\mu+iz(E)] - 2\arg\Gamma[a+iz(E)]. \tag{6}$$

---

[†] We assume that the function $E(z^2)$ is invertible giving $z^2(E)$ and that $z(E) > 0$ for $E \in \mathcal{X}$.

The corresponding total wavefunction (1) is determined once the continuous and discrete Fourier components are given as shown by Eq. (2). Therefore, having decided on the energy spectral function $E(z^2)$ and chosen $P_n(z^2)$ as $S_n^\mu(z^2;a,a)$, we only need for this purpose the basis set $\{\phi_n(x)\}$ and the continuous and discrete weight functions $\rho(z)$ and $\omega_k$. These weight functions are given in Section IV.B of Ref. [1] as follows

$$\rho(z) = \frac{\left|\Gamma[\mu+iz(E)]/\Gamma[2iz(E)]\right|^2 \Gamma[a+iz(E)]\Gamma[a-iz(E)]}{2\pi\,\Gamma(2a)\Gamma^2(\mu+a)}, \tag{7}$$

$$\omega_k = \frac{(k+\mu)}{k!}\frac{2(-1)^{k+1}(2\mu)_k}{\Gamma(2a)\Gamma(1-2\mu)}\left[\frac{\Gamma(a-\mu)(\mu+a)_k}{(\mu-a+1)_k}\right]^2, \tag{8}$$

where $(a)_n = a(a+1)(a+2)...(a+n-1) = \frac{\Gamma(n+a)}{\Gamma(a)}$ is the Pochhammer symbol (the shifted factorial). Therefore, for a given energy spectral function $E(z^2)$ and basis $\{\phi_n(x)\}$, full analytic determination of the objects in equations (2), (5), (6), (7) and (8) are obtained. That is, we end up with the following analytically realizable components of the system: (i) energy spectrum, (ii) scattering phase shift, and (iii) wavefunction. Hence, we will not ponder any more on these three items but will seek the corresponding potential function that can only be determined numerically for a given $E(z^2)$, $\{\phi_n(x)\}$ and a given set of physical parameters.

## 3. The potential matrix

The one-dimensional Schrödinger wave equation is $i\frac{d}{dt}\Psi(x,t) = H\Psi(x,t)$, where $H$ is the Hamiltonian operator. Substituting (1) and (2), in this equation then projecting from left by $\langle\phi_m(x)|$ and integrating over $x$ gives the following equivalent two matrix wave equations

$$\sum_n \mathcal{H}_{m,n} P_n(z^2) = E\sum_n \Omega_{m,n} P_n(z^2), \tag{9a}$$

$$\sum_n \mathcal{H}_{m,n} P_n(z_k^2) = E_k \sum_n \Omega_{m,n} P_n(z_k^2), \tag{9b}$$

where $\mathcal{H}_{m,n} = \langle\phi_m|H|\phi_n\rangle$ and $\Omega_{m,n} = \langle\phi_m|\phi_n\rangle$. On the other hand, the three-term recursion relation (3b) could be rewritten as

$$\sum_n R_{m,n} P_n(z^2) = z^2 \sum_n \delta_{m,n} P_n(z^2), \tag{10}$$

where $R$ is the tridiagonal symmetric matrix whose elements are: $R_{m,n} = A_m \delta_{m,n} + B_{m-1} \delta_{m,n+1} + B_m \delta_{m,n-1}$. To simplify the analysis, we choose the basis $\{\phi_n(x)\}$ as an orthonormal set making $\Omega_{m,n} = \delta_{m,n}$. Thus, we can write Eq. (9) and Eq. (10) as $\mathcal{H}|P\rangle = E|P\rangle$ and $R|P\rangle = z^2|P\rangle$, respectively. In other words, $|P\rangle$ is a common eigenvector for the two Hermitian matrices $\mathcal{H}$ and $R$ with the corresponding eigenvalues $E$ and $z^2$. Now, since $E$ is a function of $z$ as $E(z^2)$, then we can write [12]

$$\mathcal{H} = \Lambda [E(R)] \Lambda^{-1}, \tag{11}$$

where $\Lambda$ is a similarity transformation (usually, an involutory matrix: $\Lambda = \Lambda^{-1} = \Lambda^T$). One such involutory matrix is the Householder transformation matrix that makes $\mathcal{H}$ a tridiagonal symmetric matrix (see, section 11.3 of Ref. [13]). Consequently, our solution is unique modulo a similarity transformation. In fact, none-uniqueness is a well-known feature of the solution of the inverse problem [14-16]. Since we have decided to choose the basis set $\{\phi_n(x)\}$ as orthonormal then we could take $\Lambda$ to be the identity matrix giving $\mathcal{H} = E(R)$ [‡].

With the Hamiltonian matrix $\mathcal{H}$ being determined, we only need the matrix representation of the kinetic energy operator $T$ in the basis $\{\phi_n(x)\}$ to obtain the potential energy matrix as $\mathcal{V} = \mathcal{H} - \mathcal{T}$. Now, in one dimension, $T$ is just $-\frac{1}{2}\frac{d^2}{dx^2}$ and in two/three dimensions with cylindrical/spherical symmetry $T = -\frac{1}{2}\frac{d^2}{dr^2} + \frac{L^2 - \frac{1}{4}}{2r^2}$ where $r$ is the radial coordinates and in 2D $L = 0, \pm 1, \pm 2, ...$ whereas in 3D $L = \ell + \frac{1}{2} = \frac{1}{2}, \frac{3}{2}, \frac{5}{2}, ...$. Therefore, it is straightforward to perform the needed differentiation and integration to obtain the matrix elements of $T$ as $\mathcal{T}_{m,n} = \langle \phi_m | T | \phi_n \rangle$. Having finally obtained the matrix elements of the potential $\mathcal{V}_{m,n}$ and the basis $\{\phi_n(x)\}$ in which they were calculated, we can use any one of four procedures introduced in Section 3 of Ref. [4] to calculate the potential function $V(x)$ locally for a given set of physical parameters. In the following section, we present several examples where we start by choosing the energy spectral function $E(z^2)$ and the orthonormal basis set $\{\phi_n(x)\}$. To validate our technique, we choose the first example to correspond to an exactly solvable and well-known system; the Morse potential.

---

[‡] Let $\{\Sigma_{m,n}\}_{m=0}^{J-1}$ be the normalized eigenvector of the $J \times J$ finite submatrix of $R$ corresponding to the eigenvalue $\lambda_n$, then one can show that $E(R) = \Sigma W \Sigma^T$ where $W$ is a diagonal matrix whose elements are: $W_{n,m} = \delta_{n,m} E(\lambda_n)$.

# 4. The potential function

In this section, we start by a choice of an *invertible* spectral function $E(z^2)$ and then employ the procedure outlined in Section 3 above to obtain the matrix representation of the potential $\mathcal{V}$ in a given basis $\{\phi_n(x)\}$ then use one or more of the four methods in Ref. [4] to derive the potential function $V(x)$. We should note that the potential function obtained as such is valid only locally. That is, the numerical plots obtained as result of this procedure may not be valid globally over the whole configuration space. To validate the accuracy of the procedure, we start by the exactly solvable system corresponding to the Morse potential. The symmetric three-term recursion relation for $S_n^\mu(z^2;a,b)$ is given by Eq. (25) in Ref. [1]. Consequently, we obtain the following matrix elements of $R$

$$R_{n,n} = A_n = (n+\mu+a)^2 + n(n+2a-1) - \mu^2, \tag{12a}$$

$$R_{n,n+1} = R_{n+1,n} = B_n = -(n+\mu+a)\sqrt{(n+1)(n+2a)}. \tag{12b}$$

## 4.1 The Morse potential

For this system, $E(z^2) = \frac{1}{2}\lambda^2 z^2$ and $\phi_n(x) = C_n y^\nu e^{-y/2} L_n^{2\nu-1}(y)$ where $\lambda$ is a positive scale parameter of inverse length dimension, $y = e^{-\lambda x}$, $L_n^{2\nu-1}(y)$ is the Laguerre polynomial with $\nu > 0$ and $C_n$ is a normalization constant chosen to make the basis set orthonormal (with respect to the integration measure $dx$) as $C_n = \sqrt{\lambda n!/\Gamma(n+2\nu)}$.

The energy spectrum is $E_k = -\frac{1}{2}\lambda^2(k+\mu)^2$ where $k = 0,1,..,\lfloor -\mu \rfloor$ and the scattering phase shift is obtained from Eq. (6) as

$$\delta(E) = \arg\Gamma(2i\kappa/\lambda) - \arg\Gamma[\mu+i(\kappa/\lambda)] - 2\arg\Gamma[a+i(\kappa/\lambda)], \tag{13}$$

where $\kappa = \sqrt{2E}$. It is well known that this energy spectrum and phase shift are associated with the 1D Morse potential $V(x) = \frac{\lambda^2}{8}\left[e^{-2\lambda x} + 2(2\mu-1)e^{-\lambda x}\right]$ [17]. We will show next that the potential function obtained numerically as outlined above does indeed reproduce this exact result if we choose $\nu = a$.

Using the differential equation, recursion relation, and orthogonality of the Laguerre polynomials, we obtain the following matrix elements of the kinetic energy operator

$$\frac{-2}{\lambda^2}\mathcal{T}_{n,m} = \frac{1}{\lambda^2}\langle\phi_n|\frac{d^2}{dx^2}|\phi_m\rangle = \frac{1}{4}(J^2)_{n,m} - \left[2(n+\nu)^2 + \nu(1-\nu)\right]\delta_{n,m}$$
$$+ \left(n+\nu-\frac{1}{2}\right)\sqrt{n(n+2\nu-1)}\,\delta_{n,m+1} + \left(n+\nu+\frac{1}{2}\right)\sqrt{(n+1)(n+2\nu)}\,\delta_{n,m-1} \tag{14}$$

where $J_{n,m} = 2(n+\nu)\delta_{n,m} - \sqrt{n(n+2\nu-1)}\,\delta_{n,m+1} - \sqrt{(n+1)(n+2\nu)}\,\delta_{n,m-1}$. The Hamiltonian matrix, on the other hand, is obtained simply as $\mathcal{H} = E(R) = \frac{1}{2}\lambda^2 R$ giving the potential matrix in the basis $\{\phi_n(x)\}$ as $\mathcal{V} = \mathcal{H} - \mathcal{T}$. Figure 1 is a plot of the potential function for a given set of physical parameters and with $\nu = a$. The figure shows an excellent match with the exact result.

All four methods in Section 3 of Ref. [4] produced identical plots for the same parameters. However, see Section 5 below for a relevant computational note.

**4.2 First non-conventional system**

This system is in 3D with spherical symmetry and we choose $E(z^2) = \frac{1}{2}\lambda^2\left(z^2 - \alpha^2 z^{-2}\right)$, $\phi_n(r) = \sqrt{\frac{2\lambda(n!)}{\Gamma(n+\ell+\frac{3}{2})}} (\lambda r)^{\ell+1} e^{-\lambda^2 r^2/2} L_n^{\ell+\frac{1}{2}}(\lambda^2 r^2)$ where $\ell$ is the angular momentum quantum number. In this case, the energy spectrum is

$$E_k = -\tfrac{1}{2}\lambda^2 \left[(k+\mu)^2 - \left(\tfrac{\alpha}{k+\mu}\right)^2\right], \tag{15}$$

where $k = 0,1,...,\lfloor -\mu \rfloor$ and $\mu$ is not an integer. The scattering phase shift is obtained using Eq. (6) with $z(E) = \frac{\sqrt{E}}{\lambda}\left[1 + \sqrt{1+(\alpha\lambda^2/E)^2}\right]^{1/2}$.

Using the differential equation, recursion relation, and orthogonality of the Laguerre polynomials, we obtain the following matrix elements of the kinetic energy operator

$$\begin{aligned}\frac{2}{\lambda^2}\mathcal{T}_{n,m} &= \frac{2}{\lambda^2}\langle\phi_n|\left[-\tfrac{1}{2}\tfrac{d^2}{dr^2} + \tfrac{\ell(\ell+1)}{2r^2}\right]|\phi_m\rangle \\ &= \left(2n+\ell+\tfrac{3}{2}\right)\delta_{n,m} + \sqrt{n\left(n+\ell+\tfrac{1}{2}\right)}\delta_{n,m+1} + \sqrt{(n+1)\left(n+\ell+\tfrac{3}{2}\right)}\delta_{n,m-1}\end{aligned} \tag{16}$$

The Hamiltonian matrix, on the other hand, is obtained as $\mathcal{H} = E(R) = \frac{1}{2}\lambda^2\left(R - \alpha^2 R^{-1}\right)$. To compute an $N \times N$ finite version of this matrix, we can use results from [18,19] to obtain the inverse of the tridiagonal symmetric submatrix of $R$ as

$$\left(R^{-1}\right)_{j,j} = \frac{C_{j+1}C_{j+2}...C_{N-1}}{D_j D_{j+1}...D_{N-1}}, \quad j = 0,1,...,N-1 \tag{17a}$$

$$\left(R^{-1}\right)_{n,m} = (-1)^{n+m} \frac{C_{m+1}C_{m+2}...C_{N-1}}{D_n D_{n+1}...D_{N-1}} B_n B_{n+1}...B_{m-1}, \tag{17b}$$

where $n = 0,1,...,N-2$ and $m = n+1, n+2,...,N-1$. The numbers $\{C_n, D_n\}$ are obtained recursively as follows

$$C_{N-1} = A_{N-1}, \quad C_n = A_n - \frac{B_n^2}{C_{n+1}}, \quad n = N-2, N-3,...,0 \tag{18a}$$

$$D_0 = A_0, \quad D_n = A_n - \frac{B_{n-1}^2}{D_{n-1}}, \quad n = 1,2,...,N-1 \tag{18b}$$

Finally, the $N \times N$ potential matrix is obtained as $\mathcal{V} = \mathcal{H} - \mathcal{T}$. The potential function $V(r)$ is obtained using these matrix elements and the basis $\{\phi_n(r)\}$. Figure 2 shows this potential function (in units of $\lambda^2$) computed using the method explained in the Appendix for a given set of physical parameters $\{\ell, \mu, a\}$ and with $\alpha = \ell/2$.

### 4.3 Second non-conventional system

For this system, we choose $E(z^2) = \frac{1}{2}\lambda^2\left(e^{\alpha z^2} - 1\right)$ with $\alpha > 0$ and $\phi_n(x) = \frac{\sqrt{\lambda}}{\sqrt{2^n n!\sqrt{\pi}}} e^{-\lambda^2 x^2/2}$ $\times H_n(\lambda x)$. In this case, the energy spectrum becomes the following remarkable formula

$$E_k = \frac{1}{2}\lambda^2\left[e^{-\alpha(k+\mu)^2} - 1\right], \tag{19}$$

where $k = 0, 1, ..., \lfloor -\mu \rfloor$. The scattering phase shift is obtained using Eq. (6) with $z(E) = \sqrt{\alpha^{-1}\ln[1 + (2E/\lambda^2)]}$.

Using the differential equation, recursion relation, and orthogonality of the Hermite polynomials, we obtain the following matrix elements of the kinetic energy operator

$$T_{n,m} = -\frac{1}{2}\langle\phi_n|\frac{d^2}{dx^2}|\phi_m\rangle = \frac{\lambda^2}{4}\left[(2n+1)\delta_{n,m} - \sqrt{n(n-1)}\delta_{n,m+2} - \sqrt{(n+1)(n+2)}\delta_{n,m-2}\right]. \tag{20}$$

The Hamiltonian matrix, on the other hand, is obtained as $\mathcal{H} = E(R) = \frac{1}{2}\lambda^2\left(e^{\alpha R} - 1\right)$ giving the potential matrix in the basis $\{\phi_n(x)\}$ as $\mathcal{V} = \mathcal{H} - \mathcal{T}$. Figure 3 shows the potential function $V(x)$ in units of $\lambda^2$ computed using the method explained in the Appendix for a given set of physical parameters $\{\alpha, \mu, a\}$.

### 4.4 Third non-conventional system

For this system, we choose $E(z^2) = \frac{1}{2}\lambda^2 \sinh(\alpha z^2)$ where $\alpha > 0$ and $\phi_n(x) = 2^\nu \Gamma(\nu)$ $\times\sqrt{\frac{\lambda(n+\nu)n!}{2\pi\Gamma(n+2\nu)}}(1-y^2)^{\frac{\nu}{2}+\frac{1}{4}} C_n^\nu(y)$, $y = \tanh(\lambda x)$, and $C_n^\nu(y)$ is the Gegenbauer (ultra-spherical) polynomial with $\nu > -\frac{1}{2}$. In this case, the energy spectrum becomes

$$E_k = -\frac{1}{2}\lambda^2 \sinh\left[\alpha(k+\mu)^2\right], \tag{21}$$

where $k = 0, 1, ..., \lfloor -\mu \rfloor$. The scattering phase shift is obtained using Eq. (6) with $z(E) = \sqrt{\alpha^{-1}\sinh^{-1}(2E/\lambda^2)}$.

Using the differential equation, recursion relation, and orthogonality of the Gegenbauer polynomials, we obtain the following matrix elements of the kinetic energy operator

$$\frac{2}{\lambda^2}T_{m,n} = \frac{-1}{\lambda^2}\langle\phi_m|\frac{d^2}{dx^2}|\phi_n\rangle = \left[n^2 + (2n+1)\nu + \frac{1}{2}\right]\delta_{m,n}$$
$$-2nG_n K_{m,n+1} + 2(n+2\nu)G_{n-1}K_{m,n-1} - \left[(n+\nu)^2 + 2\nu + \frac{3}{4}\right](K^2)_{m,n} \tag{22}$$

where $K_{n,m} = G_{n-1}\delta_{n,m+1} + G_n\delta_{n,m-1}$, $G_n = \frac{1}{2}\sqrt{\frac{(n+1)(n+2\nu)}{(n+\nu)(n+\nu+1)}}$ and we have used the differential property

$$\left(1-y^{2}\right)\frac{d}{dy}C_{n}^{\nu}(y)=\frac{1/2}{n+\nu}\left[(n+2\nu)(n+2\nu-1)C_{n-1}^{\nu}(y)-n(n+1)C_{n+1}^{\nu}(y)\right]. \quad (23)$$

The Hamiltonian matrix, on the other hand, is obtained as $\mathcal{H}=E(R)=\frac{1}{2}\lambda^{2}\sinh(\alpha R)$ giving the potential matrix in the basis $\{\phi_n(x)\}$ as $\mathcal{V}=\mathcal{H}-\mathcal{T}$. Figure 4 shows the potential function $V(x)$ in units of $\lambda^2$ computed using the method explained in the Appendix for a given set of physical parameters $\{\alpha,\nu,\mu,a\}$.

## 5. Comment on the computational procedures in Sec. 3 of Ref. [4]

In some problems, one or more of the four procedures in Section 3 of Ref. [4], produce artificial oscillations in the potential function plot that increases with the size of the potential matrix. However, the oscillating curve intersects the correct potential function curve only at $N$ points $\{x_n\}_{n=0}^{N-1}$ where $N$ is the size of the potential matrix $\mathcal{V}$. The points $\{x_n\}_{n=0}^{N-1}$ where the potential function assumes the correct values $\{V(x_n)\}_{n=0}^{N-1}$ can be found for a basis set $\{\phi_n(x)\}$ that could be written as $\phi_n(x)=W(y)Q_n(y)$ where $y=g(x)$, $W(y)$ is some positive weight function, and $\{Q_n(y)\}$ is a set of orthogonal polynomials in $y$ that satisfy the following symmetric three-term recursion relation

$$yQ_n(y)=a_nQ_n(y)+b_nQ_{n+1}(y)+b_{n-1}Q_{n-1}(y). \quad (24)$$

We construct the $N\times N$ symmetric tridiagonal matrix (the associated Jacobi matrix) as $J_{n,m}=a_n\delta_{n,m}+b_{n-1}\delta_{n,m+1}+b_n\delta_{n,m-1}$. Thus, Eq. (24) becomes an eigenvalue equation $J|Q\rangle=y|Q\rangle$ and the corresponding eigenvalues $\{y_n\}_{n=0}^{N-1}$ give $\{x_n\}_{n=0}^{N-1}$ as solutions of $y_n=g(x_n)$. Then, we use a conveniently chosen fitting routine to fit the $N$ pair of points $\{x_n,V(x_n)\}_{n=0}^{N-1}$ to a smooth curve $V(x)$. The curve accuracy increases with the size $N$ of the potential matrix $\mathcal{V}$. In this work, we used the continued fraction fitting routine based on the rational fraction approximation of Haymaker and Schlessinger similar to that in the Padé method [20].

## Appendix

In this appendix, we detail one of the methods to calculate the potential function using its matrix representation in a given basis set. Let $V$ denotes the quantum mechanical Hermitian operator that stands for the real potential energy and let $\mathcal{V}$ be its matrix representation in a given square-integrable basis set $\{\phi_n(x)\}_{n=0}^{\infty}$. That is, $\mathcal{V}_{n,m}=\langle\phi_n|V|\phi_m\rangle=\int\phi_n(x)V(x)\phi_m(x)dx$. Using Dirac notation, we can write

$$\langle x|V|x'\rangle=V(x)\delta(x-x'), \quad (A1)$$

where $\delta(x-x') = \langle x | x' \rangle$ and where $x$ stands for the configuration space coordinate. Moreover, $\langle x | \phi_n \rangle = \phi_n(x)$ and $\langle x | \bar{\phi}_n \rangle = \bar{\phi}_n(x)$ where $\langle \phi_n | \bar{\phi}_m \rangle = \langle \bar{\phi}_n | \phi_m \rangle = \delta_{n,m}$. Using the completeness of configuration space, $\int |x\rangle\langle x| dx = 1$, we can write the expression $\langle x | V | \phi_n \rangle$ as follows

$$\langle x | V | \phi_n \rangle = \int \langle x | V | x' \rangle \langle x' | \phi_n \rangle dx' = V(x)\phi_n(x), \tag{A2}$$

where we have used (A1) in the last step. The completeness of the basis reads $\sum_m |\bar{\phi}_m\rangle\langle\phi_m| = \sum_m |\phi_m\rangle\langle\bar{\phi}_m| = I$, where $I$ is the identity. Using this completeness enables us to write the left side of Eq. (A2) as

$$\langle x | V | \phi_n \rangle = \sum_{m=0}^{\infty} \langle x | \bar{\phi}_m \rangle \langle \phi_m | V | \phi_n \rangle = \sum_{m=0}^{\infty} \bar{\phi}_m(x) \mathcal{V}_{m,n}. \tag{A3}$$

Equations (A2) and (A3) give the potential function as $V(x) = [\phi_n(x)]^{-1} \sum_{m=0}^{\infty} \bar{\phi}_m(x) \mathcal{V}_{m,n}$. For finite $N \times N$ matrix calculation, we obtain the following approximation

$$V(x) \cong \frac{1}{\phi_n(x)} \sum_{m=0}^{N-1} \bar{\phi}_m(x) \mathcal{V}_{m,n}, \quad n = 0, 1, ..., N-1. \tag{A4}$$

Therefore, we need the information in only one column of the potential matrix (or one row, since $\mathcal{V}_{m,n} = \mathcal{V}_{n,m}$) to determine $V(x)$. In particular, if we choose $n = 0$, we obtain

$$V(x) \cong \frac{1}{\phi_0(x)} \sum_{m=0}^{N-1} \bar{\phi}_m(x) \mathcal{V}_{m,0}. \tag{A5}$$

In all four subsections of Section 4, the basis elements are orthonormal ($\bar{\phi}_n = \phi_n$) and could be written as $\phi_n(x) = W(y) Q_n(y)$ where $y = y(x)$ is a coordinate transformation and $W(y)$ is a positive weight function. Moreover, $Q_n(y)$ is an orthogonal polynomial of degree $n$ in $y$ with $Q_0 = 1$. Therefore, Eq. (A5) becomes

$$V(x) \cong \sum_{m=0}^{N-1} Q_m(y(x)) \mathcal{V}_{m,0}. \tag{A6}$$

# Figure Captions

**Fig. 1**: The Morse potential function shown as a solid red trace and calculated using its matrix representation in the basis of Sec. 4.1. The exact potential function shown as dotted blue trace. The energy spectrum is shown by the horizontal solid lines. The energy and potential are measured in units of $\lambda^2$. The physical parameter values are chosen as $\mu = -3.7$ and $a = 2.5$ whereas the basis parameter $\nu = a$.

**Fig. 2**: The potential function of Sec. 4.2 calculated using its matrix representation in the basis given therein. It is shown as a solid black trace whereas the energy spectrum is shown as horizontal dotted lines. The energy and potential are measured in units of $\lambda^2$. The physical parameter values are $\{\mu, a, \ell, \alpha\} = \{-7.7, -\mu, 1, \frac{1}{2}\ell\}$.

**Fig. 3**: The potential function of Sec. 4.3 calculated using its matrix representation in the basis given therein. It is shown as a solid black trace whereas the energy spectrum is shown as horizontal dotted lines. The energy and potential are measured in units of $\lambda^2$. The physical parameter values are $\{\mu, a, \alpha\} = \{-4.3, -\mu, 0.2\}$.

**Fig. 4**: The potential function of Sec. 4.4 calculated using its matrix representation in the basis given therein. It is shown as a solid black trace whereas the energy spectrum is shown as horizontal dotted lines. The energy and potential are measured in units of $\lambda^2$. The physical parameter values are $\{\mu, a, \alpha\} = \{-3.2, -\mu, 0.3\}$ whereas the basis parameter $\nu = a$.

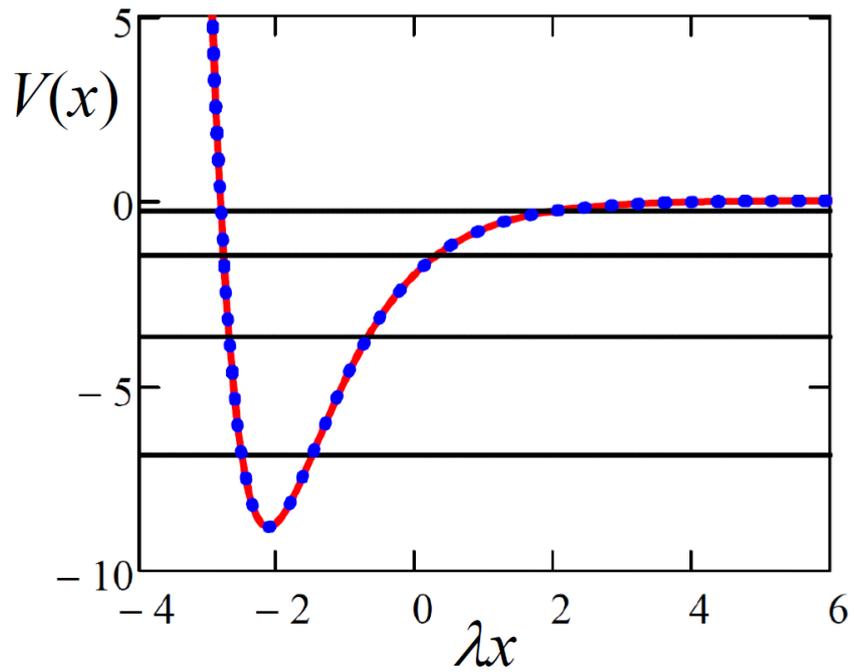

**Fig. 1**

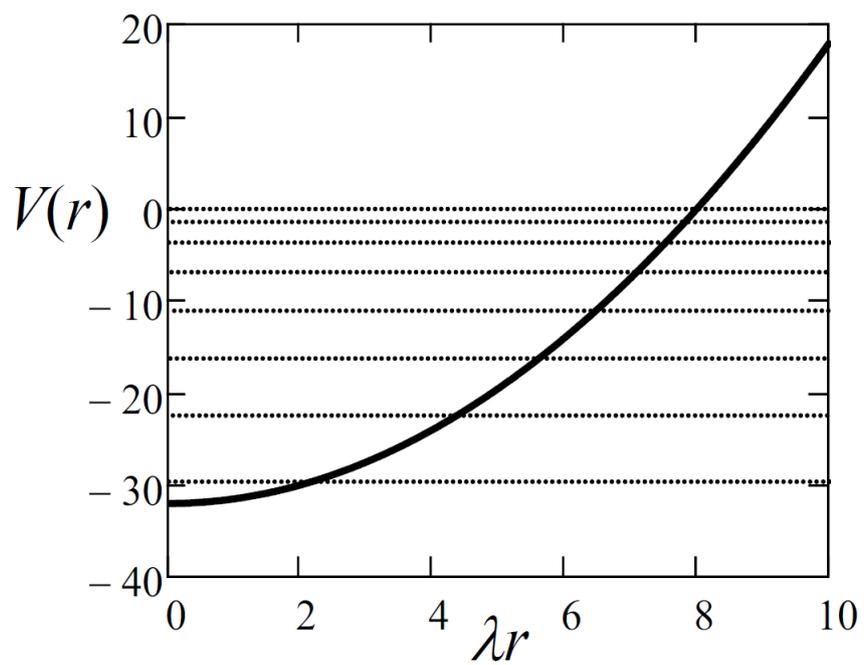

**Fig. 2**

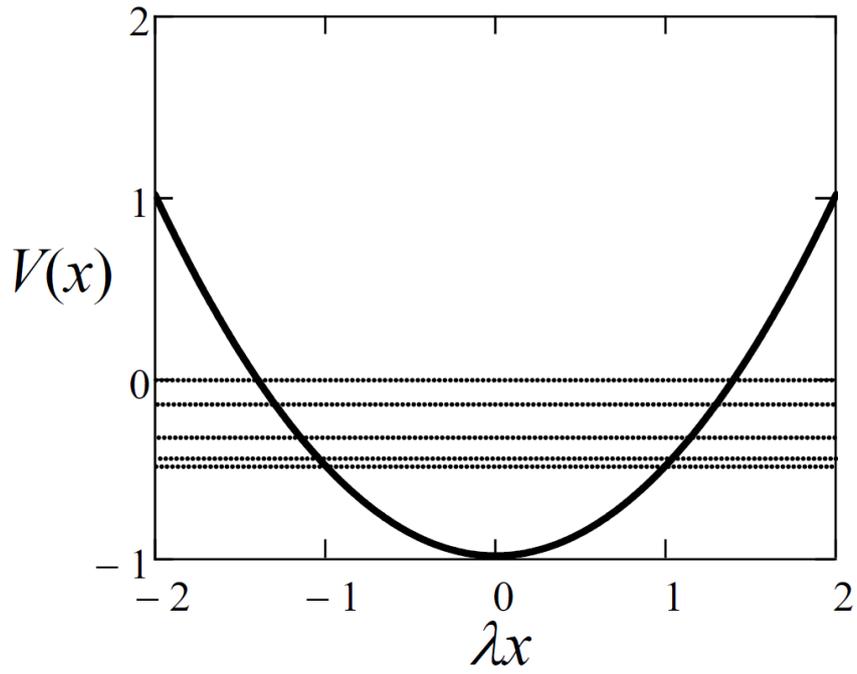

**Fig. 3**

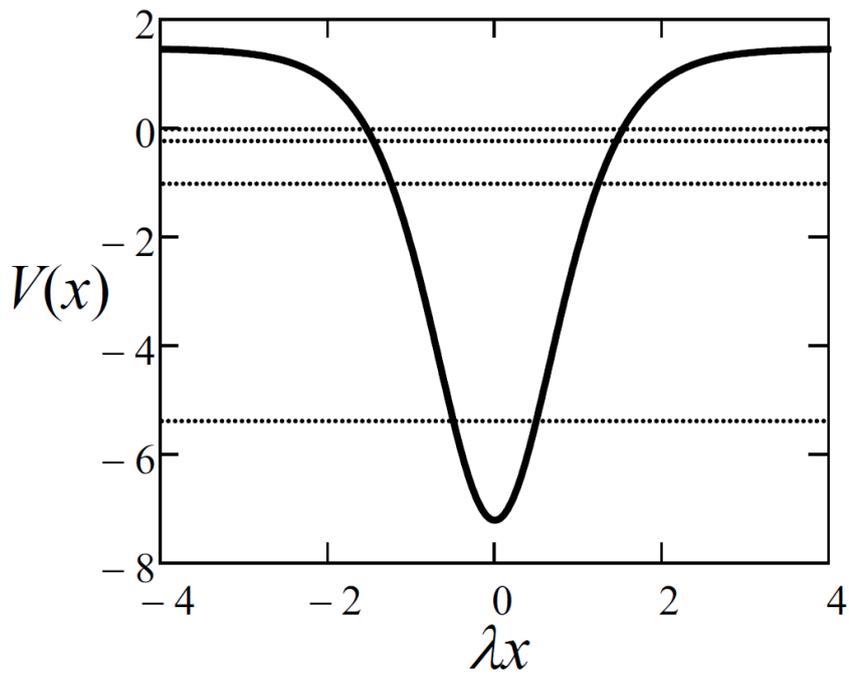

**Fig. 4**